\input preprint.sty
%
%
%
%
\title{Series studies of the Potts model. I:
The simple cubic Ising model}

\author{A J Guttmann\dag\ and I G Enting\ddag}

\address{\dag Department  of Mathematics
 The University of  Melbourne, Parkville, Victoria, Australia 3052}

\address{\ddag CSIRO, Division of Atmospheric Research,
Private Bag No. 1, Mordialloc, Victoria, Australia 3195}

\jnl{\JPA}

\shorttitle{Simple cubic Ising series}

\beginabstract
The finite lattice method of series expansion is generalised to the
$q$-state Potts model on the simple cubic lattice.
 It is found that the computational effort grows
exponentially with the square of the number of series terms obtained,
unlike two-dimensional lattices where
 the computational requirements grow  exponentially with
the number of terms.  For the Ising ($q=2$) case  we have extended
low-temperature series for the partition functions,
 magnetisation and zero-field susceptibility to
$u^{26}$ from $u^{20}$.
The high-temperature series for the zero-field partition
function is  extended from $v^{18}$ to $v^{22}$.
Subsequent analysis gives critical exponents in agreement
with those from  field theory.
\endabstract

\section{Introduction}

This is the first of a series of papers describing the application
of the finite lattice method of series expansion to various
cases of the $q$-state Potts model. A major objective in this work
is the study of the $q=3$ case in three dimensions because of
its relevance to quantum chromodynamics. However in the course of
this project we have developed extremely powerful techniques of series
expansion. This has made it possible to obtain considerable extensions
to many Potts model series and so we have been able to investigate
a large range of lattice statistics problems.

In this paper we describe the application of the finite lattice
method of series expansions to
the derivation of high- and low-temperature expansions for the free energy
of the Potts model on the simple cubic lattice. We present series
for the Ising model which is the $q=2$ case of the Potts model.
 We have been able to extend the
low-temperature series for the zero-field partition function,
magnetisation and susceptibilty from order
$u^{20}$  (Sykes \etal 1973) or $u^{24}$ (M.F. Sykes, unpublished)
 to $u^{26}$. The zero-field high-temperature series
for the partition function has been extended from
 $v^{18}$ (Sykes \etal 1972)
 to $v^{22}$. The variables $u = \exp(-4J/kT)$ and
 $v = \tanh(J/kT)$ are the usual low- and high-temperature
 model expansion variables for the Ising model.
 (A recent paper by Bhanot \etal~(1992) calculates low-temperature
 series for the simple cubic Ising model using a variant of the
 finite lattice method that is apparently less efficient than ours.
 They obtain series for the magnetisation to $u^{20}$ and
  the internal energy to order $u^{24}$, thus duplicating the work of
Sykes \etal (1973) and Sykes (unpublished) respectively.)

The extension of these series is particularly pertinent as their
behaviour is quite difficult to determine. While the high-temperature
 series are well-behaved, with uncertainty as to critical exponents
being confined to the fourth, or at the worst third, decimal place, the
situation at low temperatures is far less satisfactory.  For many years
there was considerable controversy as to the symmetry or otherwise of
the critical exponents above and below the critical temperature. That
this is now less controversial is not based on good numerical evidence
however, but rather on a strengthening belief in universality based
on a better understanding of the connection between field theory and
critical phenomena. Nickel (1991) has also recently argued for the
extension of perturbation series, in order to remove any lingering
doubts about the equivalence of the $\phi^{4}$ field theory, and the
lattice models.

The new low-temperature series we have obtained here permit an improved
analysis, though the quality of the numerical estimates is still
inferior to that of the high-temperature series. Nevertheless, by
biasing our analysis with the value of the critical temperature, obtained
from the high-temperature series, and by assuming a confluent exponent
around $\frac{1}{2}$ (the exact value is unimportant), exponent estimates are
obtained that are consistent with the best field theory estimates.
We have also extended the high-temperature zero-field free energy
series, which allows us to estimate $\alpha$ considerably more accurately
than has been possible previously.

The finite lattice method of series  expansions for lattice statistics
problems has been
applied to a number of different systems
 on a range of two-dimensional lattices.  The
initial application was in calculating the high-temperature expansion of
the three-state
Potts model (de Neef, 1975; de Neef and Enting, 1977).
The expansion of the limit of chromatic
polynomials (Kim and Enting, 1979) was formally a high-temperature
expansion and the enumeration of self-avoiding
polygons (Enting, 1980a; Enting and Guttmann, 1985;
Guttmann and Enting, 1988a) is closely related to high-temperature
expansions.  However
most of the subsequent applications of the method (based on the work of
Enting, 1978a)
have been in the derivation of low-temperature or high-field expansions,
generally on the square lattice (e.g.~Enting, 1980b; Adler \etal
1983).
Series for triangular lattice systems have been obtained by
regarding them as square lattices with additional interactions (Enting,
1980c; Enting and Wu, 1982).  Specific combinatorial expressions
for triangular lattice systems are
known (Enting, 1980d, 1987a). However these involve the
additional complication of calculating the partition functions of
hexagonal finite lattices. The only application using full
triangular symmetry has been in the enumeration of polygons
on the triangular lattice (Enting and Guttmann, 1992).
The finite lattice method has also been applied to Potts models
on the checkerboard lattice (Enting, 1987b) and polygons
on the honeycomb lattice (Enting and Guttmann, 1989). These
last two cases were treated as modifications of the square lattice.

Enting (1978b) quoted some of the combinatorial
results required for the finite lattice
method on the simple cubic lattice but until now
these results have not been applied.
In this paper we describe the formalism
for obtaining either high-temperature or low-temperature
expansions for the Potts model on the simple cubic lattice.
We present and analyse specific results for the Ising model case.

The outline of the remainder of the
paper is as follows. Section 2 analyses the combinatorial aspects
of applying the finite lattice method on the simple cubic lattice.
Section 3 describes the expansions that we have calculated.
Section 4 describes an analysis of these series.

\section{Finite Lattice Expansions on the Simple Cubic Lattice}

We begin by defining the notation that we shall use throughout
this series of papers. The Potts model is defined on a lattice
in terms of a `spin' variable,
$s_j$ at each site, $j$, taking integer values from
0 to $q-1$. There is  an  energy difference, $\Delta E$,
 between aligned and non-aligned states that interact. Generally
 we consider interactions confined to pairs of nearest-neighbour sites,
 i.e.~those joined by the bonds of the lattice.
   Sites aligned in the `0' direction
differ by a field energy $H$ from sites with other alignments.
Thus with $\delta(m,n) = 1$ of $m=n$ and 0 otherwise, the
Hamiltonian is written
$$
\hat{H} = \sum_{(i,j)} \Delta E (1 - \delta(s_i,s_j) )
 + \sum_i H (1- \delta(s_i,0))
\eqno(1)
$$
where the first sum is over all pairs of interacting sites
and the second sum is over all sites.

 The low-temperature expansion variable is
$z=\exp(-\beta \Delta E)$ where $\beta = 1/kT$.
For the field dependence,  we use the expansion
variables $\mu = \exp(-\beta H)$ and $x=1-\mu$.  The general
Potts model high-temperature expansion variable
is $v=(1-z)/(1+(q-1)z)$.

For the Ising model (i.e. $q=2$), only even powers of $z$
occur and the natural low-temperature expansion variable is $u=z^2
= \exp(-4\beta J)$  and the
high-temperature expansion variable can be written as
 $v =\tanh(\beta J)$ with $2J = \Delta E$.
For zero-field on bi-partite lattices, only even powers of $v$
have non-zero coeffcients in the high-temperature Ising model expansion.

The basic formulation of the finite lattice method
approximates the partition function per site, $Z$, as
$$
 Z = \lim_{|\Gamma| \rightarrow \infty} Z^{1/|\Gamma|}_\Gamma \approx
\prod_{\alpha \in A} Z^{W(\alpha)}_{\alpha}
\eqno(2a)
$$
where $\Gamma$ denotes a lattice that becomes arbitrarily large
in all directions and $|\Gamma|$ denotes the number of sites
in $\Gamma$.
Here $A$ is a set of finite lattices, $\alpha$, with $A$ closed
under the operation of intersection of finite lattices.
For the simple cubic lattice, this general relation has the
specific form:
$$
 Z = \lim_{N \rightarrow \infty} Z^{1/N^{3}}_{NNN} \approx
\prod_{[p,q,r] \in A} Z^{W(p,q,r)}_{pqr}
\eqno(2b) $$
where $Z_{pqr}$ is the partition function of a cuboid of dimensions
 $p \times q \times
r$ sites.  For low-temperature expansions, the $Z_{pqr}$ are to be
evaluated with a surrounding layer of fully-ordered sites.
 For high-temperature expansions, the
$Z_{pqr}$ are to be evaluated with free boundary conditions.
(It is also convenient to remove common factors as described below.)
  The weights $W(p,q,r)$
depend on the set, $A$, over which the product is taken.  In
approximations (2a, 2b) an appropriate choice of weights
 will give $Z$ as a series correct up to, but not
including, the order of the first connected graph that will not fit into
any of the cuboids of set $A$ (Enting, 1978a).

The method of Bhanot \etal (1992) uses  generalised helical
boundary conditions imposed in a sequence of configurations so that
they can ultimately remove the effect of unwanted graphs that
occur under helical or periodic boundary conditions.

Inspection of the low-temperature expansion of the Potts model shows that
the limiting
graphs are trees that do not double back in any direction:  all planes
drawn perpendicular to bonds of the
 lattice intersect such trees at most once.  Such a tree
can span a cuboid of size $p \times q \times r$ with $p+q+r-2$ sites and
$p+q+r-3$
bonds in the tree and will give powers of $4(p+q+r)-6$ or more in the Potts
model low-temperature variable, $z$.

Therefore it is appropriate to take $A$ as the set of cuboids whose
 spans, $\sigma$, satisfy
$$
 \sigma = p+q+r \leq s
 $$
For this choice of $A$,
 the first incorrect low-temperature term is of order $4(s+1)-6$,
i.e. the series
is correct to order $4s-3$.  We use the notation $A(s) = \{[p,q,r]:p+q+r
\leq s\}$ to denote the set of cuboids used.

Inspection of the high-temperature expansion shows that,
for general fields, the same tree
graphs are limiting.  However, for zero field, the limiting graphs are
maximally-extended
polygons.  These are polygons that have at most two intersections with any
plane perpendicular to the lattice bonds.
 These are the three-dimensional generalisation of
the convex polygons (Guttmann and Enting, 1988b)
that were used to obtain corrections
in the polygon enumeration by Guttmann
and Enting (1988a).  For cuboids of span $s+1$ the
maximally-extended polygons will have $2(s+1)-6$ steps so that cuboids of
span $\leq s$
will give the zero-field high-temperature series correct to $2s-5$.  The
combinatorial factors from Enting (1978b) give:

$$ W(d,w,\ell) = \sum_{[p,q,r] \in A(s)} \eta(p-d) \, \eta(q-w)
\,\eta(r-\ell)
\qquad\hbox{{\rm for} $[d,w,L] \in A(s)$}
\eqno{(3)}$$
where
$$\eqalignno{\eta(0)   &=   1  &(4a) \cr
\eta(1)   &=   -2  &(4b) \cr
\eta(2)   &=   1 &(4c) \cr
\eta(k)   &=   0 \qquad\hbox{\rm otherwise.}&(4d)\cr} $$
This implies
$$\eqalignno{ W(d,w,\ell)   &=   1 \qquad\hbox{for  $ d+w+\ell   =   s$}
&(5a)\cr
   &=   -5 \qquad\hbox{for  $d+w+\ell   =   s-1$} &(5b)\cr
   &=   10 \qquad\hbox{for  $d+w+\ell   =   s-2 $} &(5c)\cr
   &=   -10 \qquad\hbox{for  $ d+w+\ell   =   s-3$}&(5d) \cr
   &=   5   \qquad\hbox{for $d+w+\ell   =   s-4 $}&(5e)\cr
   &=   -1  \qquad\hbox{for $d+w+\ell   =   s-5 $}&(5f) \cr
   &=   0   \qquad\hbox{\rm otherwise.} &(4g) \cr}  $$
In actual computation it is convenient to exploit the cubic symmetry and
consider only
$d \leq w \leq \ell$.  We define $B(s) = \{[p,q,r]:p+q+r \leq s, p \leq q
\leq r \}$.
The expansion becomes
$$Z \approx \prod_{[p,q,r] \in B(s)}Z_{pqr}^{V(p,q,r)}
\eqno(6a)$$
with
$$
V(p,q,r) = \sum_{\pi(p,q,r)}W(\pi(p,q,r))
\qquad\hbox{for $p \leq q \leq r$}
\eqno(6b)$$
where the sum is over all {\it distinct} permutations, $\pi(p,q,r)$
of the indices.

The partition functions are constructed by using a transfer-matrix
formalism to build up
$\ell$ layers of size $d \times w$.  As in all of the most recent
applications of the finite lattice method, we used
the approach of building up the finite lattices
one site at a time.  The computational complexity of the calculation is
determined by
the value of $d \times w$ that is required and this is determined by the
limiting span $s$.
Exploiting the cubic symmetry allows the series to be generated using
cuboids whose size is subject to:

$$1   \leq   d   \leq   d_{\rm max} \quad\hbox{and}\quad
d   \leq   w   \leq   \lfloor(s-d)/2 \rfloor \quad \hbox{and}\quad
w   \leq   \ell   \leq   s-d-w$$

If $s=3m+1$ then $d_{\rm max} = m$
and the maximum size of $d \times w$ is $m \times
m$.  If $s=3m-1$ then $d_{\rm max} = m-1$
 and the largest size of $d \times w$ is $(m-1)
\times m.$  (The case of $s=3m$ has $d \times w = m \times m$
 but this is of little interest because the transfer matrices
 will be the same size as is required for the case $s=3m+1$
which gives 4 extra low-temperature series terms and 2 extra
high-temperature terms.)

The limiting factor in the finite
lattice series computations is the size of
the vector required to store the site
configurations at the end of a partly-built lattice.
In a departure from the earlier high-temperature calculations noted in the
introduction, it has been desirable to
 use a site representation rather than a bond representation
when calculating the $Z_{pqr}$ required for
high-temperature expansions. Thus size considerations are the same
for both high- and low-temperature expansions.
 For the
expansions of $q$-state models on the simple cubic lattice
the number of site configurations required is
$q^{k}$ with $k=d
\times w.$  For the $q$-state Potts model ($q \geq 3$) the symmetry between
the states
can be used to remove redundant elements.  There will be approximately
$q^{k}/(q-1)!$
distinct vector elements.  The precise number can be calculated using a
recurrence
relation.  Since the present paper only considers $q=2$ for which exactly
$2^k$ states
are required, the general recurrence relation will be described
in the second paper.
For zero-field high-temperature expansions, there is no need to
distinguish the `0' state and so for a cross-section of $k$ sites,
approximately $q^k/q!$ vector elements are needed.

Thus vectors of approximately $q^{m^2}/(q-1)!$ elements give Potts series
correct to $z^{12m+1}$ while vectors of $q^{m(m-1)}/(q-1)!$ elements
give Potts
series to $z^{12m-7}$.  Vectors of size $2^{m^2}$ elements give Ising
series correct to  $u^{6m}$ while vectors of size
$2^{m(m-1)}$ elements give Ising series correct to  $u^{6m-4}.$
For zero-field high-temperature expansions,
 vectors of approximately $q^{m^2}/q!$ elements give Potts series
correct to
$v^{6m-3}$ while vectors of $q^{m(m-1)}/q!$ elements give Potts
series to $v^{6m-7}$.  Vectors of size $2^{m^2-1}$ elements
give Ising series correct to $v^{6m-4}$  while vectors of size
$2^{m^2 -m-1}$ elements give Ising series correct to $v^{6m-8}$.

In contrast, for the square lattice, the cross section, $k$, corresponds to
the width of
the lattice (expressed as a number of sites) and so vectors of size
$\approx
q^{k}/(q-1)!$, will give Potts model low-temperature series correct to
$4k+3.$  Duality gives the high-temperature
 series to the same order.  Table 1 lists the low-order
values of $m$ together with the number of series terms obtainable for
models on the
simple cubic lattice and the size of vector required for the Ising model.

While
Bhanot \etal (1992)  compared the combinatorial complexity the
combinatorial complexity of their technique to the finite lattice method
with periodic boundary conditions, they made no comparisons with our
formalism (see Enting, 1987a; Guttmann and Enting, 1990b) using fixed
boundary conditions. Their note gives too few details of the
characterstics of their cancellation procedure for us to make a general
comparison. However from the specific example, it appears that they
need 24 spins (i.e.~$2^{24}$ configurations) to reach $u^{24}$, while,
as shown in Table 1, we require only 16 spins.

In principle,
it is possible to consider extending the series by determining
correction terms to the finite lattice expansion.  Previous studies
(Enting and Wu, 1982; Guttmann and
Enting, 1988b) have obtained explicit expressions for the leading-order
correction terms for various models.
For the minimal spanning trees direct enumeration seems difficult.
However examination
of the series shows that $M$ such trees will give a correction of
$$
(q-1)M \mu^{s-1}[z^{4s-2} + \frac{1}{2}(q-2)(s-2)z^{4s-1}]
$$
to the low-temperature series for the $q$-state Potts model.  For $q \geq
3$ the number
$M$ can be determined by noting the correction required at the
corresponding $s$ value
for the less complex $q=2$ (Ising model) case.  For the high-temperature
series, the zero-field correction from maximally extended-polygons
is of the  form $K(q-1)v^{2s-4}$.  Again the Ising case can be
used to determine the correction for higher $q$.  Since the smaller the
$q$ the larger
the lattice that can be used, our highest-order Ising cases can be used to
provide the correction terms for future $q=3$ studies,
as used in paper III of this series (Enting and Guttmann 1993, in preparation).

\section{Expansions}

The finite lattice method obtains the high- and low-temperature
expansions for the partition function, $Z$. For actual calculations,
specific normalisations must be chosen. For low-temperature expansions,
the appropriate choice is to define the partition function such that
the fully aligned (all sites `0') state is taken as having zero
energy, since the low-temperature series is a perturbation expansion
about this state. This is the normalisation that we have
chosen for equation (1) above, defining the Hamiltonian.
In this normalisation, the partition function
is often denoted $\Lambda$. On the simple cubic lattice,
the Ising model expansion begins
$$
\Lambda = 1 + u^3 \mu + 3 u^5 \mu^2 + \dots
\eqno(7)$$
As noted above, we express the field variable as $\mu = 1-x$ and,
in order to reduce the memory required by the computer
program, truncate the field dependence at $x^2$
so that
$$
\Lambda = \Lambda_0(u) + x \Lambda_1(u) + x^2 \Lambda_2(u) +  \dots
\eqno(8)$$
We express $\Lambda_0$ as
$$
\Lambda_0 = \sum_{n=0}^\infty \lambda_n u^n
\eqno(9)$$
and define an order parameter
$$
M = 1 + \frac{q}{q-1} \frac{\Lambda_1}{\Lambda_0}
= \sum_{n=0}^\infty m_n u^n
\eqno(10)$$
and susceptibility
$$
\chi = 2 \frac{\Lambda_2}{\Lambda_0}
-\frac{\Lambda_1}{\Lambda_0}
- \biggl(\frac{\Lambda_1}{\Lambda_0}\biggr)^2
\eqno(11)$$
We have determined the low-temperature expansion using cross-sections
up to $4 \times 5$, giving $\Lambda_0$, $M$ and $\chi$ correct
to $u^{26}$. The coefficients $\lambda_n$, $m_n$, $c_n$ are
listed in Table 2.

The calculations build up the finite lattice one site at a time,
running through all $q$ states of the site and applying a weight
of 1 for state `0' and $\mu$ (expressed as $1-x$) for all other
states. The bonds linking each new site to the partly constructed
lattice are given weight 1 if the sites at each end are in the same
state and weight $z$ otherwise. For low-temperature expansions,
the set of bonds considered includes bonds connecting the
finite lattice to an outer boundary of sites taken as being in
state `0'.

The high-temperature expansions have been obtained
using cross-sections up to $4 \times 5$, giving $Z$ to order $v^{22}$.
While table 1 indicates that
the vector size increases by a factor of 16
(or more generally $q^4$) on going from $4 \times 4$
to $4 \times 5$, for high temperature expansions, the memory
requirements are
reduced by a factor of  2 (or more generally a factor of
$q$) by making full use of the symmetry and  by a  further factor of 3
since the high-temperature field dependence is truncated at zeroth
order compared to the $x^2$ truncation of the low-temperature series.
In principle it is possible to further reduce the memory
requirements for the high-temperature Ising free energy
calculation by
making use of the ferromagnetic/antiferromagnetic
symmetry of the zero-field Ising model to reduce the vector
sizes by a further
factor of 2. Since we have been using programs designed
for general $q$, we have not exploited this possibility.

For the high-temperature expansions, the field weighting is not
included and the set of bonds does not include any
bonds  extending beyond
the finite latttice. However the main difference from the
low-temperature expansion arises from the relation
$z = (1-v)/(1 + (q-1)v)$. For a finite lattice $\alpha$, rather
than expand $\Lambda_\alpha$ we expand:
$$
\Phi_\alpha = q^{-s(\alpha)}(1+ (q-1)v)^{b(\alpha)}\, \Lambda_\alpha
\eqno(12)$$
whence
$$
\Lambda_\alpha = q^{s(\alpha)-b(\alpha)}(1+ (q-1)z)^{b(\alpha)}\, \Phi_\alpha
\eqno(13)$$
where $b(\alpha)$ and $s(\alpha)$ are the number
of bonds and sites respectively in lattice $\alpha$.
The expansion of $\Phi_\alpha$ is obtained by giving bonds
a weight of $1 + (q-1) v$ for pairs of sites in the same state and weight
$1-v$ otherwise. An additional weight of $q^{-1}$ is applied at
each site.

For the infinite lattice limit, we have
$$
q^{1-\nu/2}(1 +(q-1)z)^{\nu/2} \,\Phi(v)
= q(1 +(q-1)v)^{-\nu/2} \,\Phi(v)
= \Lambda \approx \prod_{\alpha \in B(s)} \Lambda_\alpha^{V(\alpha)}
\eqno(14)$$
where $\nu$ is the lattice coordination number so that
$\nu/2$ is the number of bonds per site,
which is 3 on the simple cubic lattice.
By using the facts that
$
\sum_{\alpha \in B(s)} b(\alpha) V(\alpha) = \nu/2
$
and
$
\sum_{\alpha \in B(s)} s(\alpha) V(\alpha) = 1
$
we obtain the relation
$$
\Phi(v) \approx \prod_{\alpha \in B(s)} \Phi_\alpha(v)^{V(\alpha)}
\eqno(15)$$
It is this last expression that we use in our
high-temperature calculations.

For the Ising model, it is usual to shift the zero of energy so that
parallel and antiparallel pairs have energies $\pm J$ with
$J = \frac{1}{2} \Delta E$. This leads to the more familiar form
of the Ising model expansion:
$$
Z = \exp(+J/kT)^{\nu/2}\, \Lambda = 2[\cosh(J/kT)]^{\nu/2}\, \Phi(v)
\eqno(16)$$

The high-temperature expansion for the free energy is written as
$$
\Phi(v) = \sum_n a_n v^n = 1 + 3v^4 + \dots
\eqno(17)$$
(or  more generally $ 1 + 3(q-1) v^4 \dots$).
The coefficients $a_n$ for $n \leq 22$ are listed in Table 2.

The internal energy, $U$, is derived from the free energy, $F$,
or the partition function $\Lambda$ by
$$
U = \frac{\partial}{\partial \beta} \beta F
= -  \frac{\partial}{\partial \beta} \ln \Lambda
$$

For low temperatures we use
$$
\frac{\d}{\d\beta} = \frac{\d z}{\d \beta}\, \frac{\d}{\d z}
= - \Delta E z \frac{\d}{\d z}
$$
to give
$$
U = \Delta E \,z\,\frac{\d \Lambda}{\d z}/\Lambda
$$

For high temperatures we use $z = (1-v)/(1 + (q-1)v)$
and $\frac{\d z}{\d v} = -q/(1+(q-1)v)^2$
to give
$$
\frac{\d}{\d \beta} = \Delta E \frac{(1-v)(1 + (q-1)v)}{q} \frac{\d}{\d v}
$$
whence
$$
U = \Delta E \frac{\nu}{2} \frac{q-1}{q} (1-v)
 - \Delta E \frac{(1-v)(1+(q-1)v)}{q} \frac{\d}{\d v} \ln \Phi(v)
 $$

The high-temperature and low-temperature expansions were
computed using two very similar Fortran programs written
for general (integer) $q$. Our programs work in terms of the
general variables $z$ or $v$, even in those special cases where
only even powers occur.
Most of the other technical aspects are similar to our previous
calculations using the finite lattice method. The calculations were
performed using arithmetic modulo various primes, $p$, slightly below
$2^{15}$. Calculations were performed using 32-bit integers and the
large vectors of residues stored as 16-bit integers. The factor
$(q-1)^{-1}$ required in the general form of $M$ is calculated as
$(q-1)^{p-2}$ modulo $p$ by virtue of Fermat's theorem.
Similarly the factors $q^{-1}$ required in evaluating the $\Phi_\alpha$
were expressed as $q^{p-2}$ modulo $p$.
Since all of the $\Phi_\alpha$ and $\Lambda_\alpha$
are of the form $1 + k v^4 + \dots$ and
  $1 + (q-1)\mu z^\nu + \dots$ respectively,
the expansions for negative powers of the $\Phi_\alpha$
and $\Lambda_\alpha$ can
readily be calculated.
The programs were run on an IBM 3090/400J with $\frac{1}{2}$ GB
of main memory and 2 GB of extended storage.

\section{Analysis}

Before presenting the analysis of our new series, we note a number of
relevant prior studies. Gaunt and Sykes (1973) extended the diamond and
f.c.c. low-temperature series, and obtained $0.307 \leq \beta \leq 0.317$
and a value for $\gamma$ around 1.27 -- 1.30. They noted that the sequences
of exponent estimates were converging rather slowly. A range of previous work
was consistent with these estimates. At high temperatures, Sykes \etal
(1972) and Camp \etal (1976) found estimates for $\alpha$ consistent
with 0.125, in accordance with the belief at the time that $\alpha$ was a
simple fraction, probably $\frac{1}{8}$.

The field theory estimates of
Le Guillou and Zinn-Justin (1980), in which scaling
and high/low-temperature exponent symmetry are implicit, are $\gamma =
1.241(2)$, $\alpha = 0.110(5)$, $\beta = 0.325(2)$. Recently Oitmaa \etal
(1991) obtained extended low-temperature series for the (2+1)
dimensional Ising model, equivalent to a $Z_2$ gauge model in (2+1)
 dimensions, which is in the three-dimensional Ising universality class.
They also derived and analysed series on the triangular and honeycomb
lattices. Unbiased exponent estimates were given as $\gamma = 1.28(2)$,
$\beta = 0.311(4)$ and $\alpha = 0.11(4)$, while biased estimates were
given as $\gamma = 1.25(2)$, $\beta = 0.318(4)$ and $\alpha = 0.096(6)$.

Nickel (1991) recently re-examined the $\phi^{4}$ field theory exponents,
and showed that, by permitting a second confluent exponent, critical
exponent estimates were obtained that agreed rather well with those
obtained from high-temperature series expansions. Nickel's preferred
values are $\gamma = 1.238$, $\nu = 0.630$, $\eta = 0.0355$. Hence $\alpha
= 0.110$ and $\beta = 0.326$. The critical exponent   $\Delta_1 =
\omega_1\nu = 0.53$.

Recently Mojumder (1991) has developed a theory based on partial
non-renormalization of superconformal dimensions of matter fields
on a (2,0) supersymmetric string world sheet. A consequence of this
theory is the predicition of critical exponents which are supposed
to be exact. These are $\alpha = \frac{1}{8}$,
$\beta = \frac{3}{8}$, from which
follows $\gamma = \frac{9}{8}$ and $\nu = \frac{5}{8}$.

Before analysing the newly obtained series, we remark that the
 analysis of the low-temperature series is made more difficult
than the analysis of their high-temperature counterparts by the
presence of non-physical singularities closer to the origin than the
physical singularity. While differential approximants effectively
afford analytic continuation beyond the nearest singularity, the
series coefficients are nevertheless dominated by the non-physical
singularity. Another approach to this difficulty is to apply a transformation,
to move the physical singularity closer to the origin than the
non-physical singularity. However,  such `singularity-moving' transformations
introduce long-period oscillations (Hunter, 1987; Guttmann 1989)
thus rendering suspect
extrapolations based on such transformed series as in the analysis by
Bhanot \etal (1992).

Another problem with the analysis of the three-dimensional Ising series
is the presence of confluent singularities (which are extremely weak
or non-existent in the two-dimensional model). Both field theory and
high-temperature series analysis suggests a value of the confluent
exponent not very different from 0.5. Roskies (1981) introduced an
effective transformation for analysing such series, in which one
replaces the original expansion variable $z$ by a new variable $y$,
defined by $1 - y = (1 - z/z_c)^{1/2}$. If the original series had
square root correction terms, the transformed series has analytic
correction terms. If there were no square root correction terms, then
nothing is lost by applying this transformation. One problem is that
the critical temperature must be accurately known, but for the
three-dimensional Ising model this is the case. The consensus of
extensive high-temperature series and Monte Carlo work is
$\tanh(J/kT_c) = 0.218093$, or $u_c = \exp(-4J/kT_c) = 0.4120494$, with
errors being a few parts in the sixth
 decimal place (Guttmann 1987b).

Our initial analysis of the three low-temperature series (specific
heat, magnetisation and susceptibility) by Pad\'e approximants and
differential approximants was not particularly illuminating. Exponent
estimates were $\alpha' = -0.1$ to $0.2$, $\beta = 0.30$ to $0.32$
and $\gamma' =
1.2$ to $1.3$. Biased estimates (with the critical point specified) were
$\alpha' \approx 0.20, \beta \approx 0.32$ and $\gamma' \approx 1.25$.

Analysing the Roskies-transformed series by evaluating Dlog Pad\'e
approximants at $y=1$ gave the results shown in Tables 3 to 5.
The approximants are fairly consistent, a few wildly deviant ones
being due to defective
approximants. For all three exponents we have taken the mean of the last
few values (from 14 to 16 approximants), and ignored the wildly different
outliers. The mean of these approximants, with error given as one
standard deviation, is
$\alpha' = 0.124 \pm 0.006$,
$\beta = 0.329 \pm 0.009$,
$\gamma' = 1.251 \pm 0.028$.

The effect of allowing for the correction-to-scaling shows
that its neglect in the analysis by Bhanot \etal (1992) is an
even more serious problem for their analysis than the use
of the Euler transformation.

It can be seen that, with the exception of the estimate of $\alpha'$, the
exponent estimates above are consistent with the field-theory estimates.
The value for $\alpha'$ just fails to overlap the range of the field
theory estimate, but we believe that the high-temperature series result
(see below) must be taken into account also.

We have also tested the sensitivity of our estimates to changes in
both the critical point estimate, and the exponent estimate used
in the Roskies transformation. Changing the estimate of the critical
point by 2 parts in the fifth significant digit produced a change
in the exponent estimate of one-tenth of the error estimates quoted
above. That change in the critical point is far greater
than the uncertainty in the critical
point, so that we can safely assert that the results quoted include
errors due to the uncertainty in the critical point. Changing the
exponent estimate to 0.53 from the value 0.5 used in the Roskies
transformation produced an error that was less than one-fifth of the
errors quoted above, so that the uncertainty in the correction-to-scaling
exponent is also negligible in this analysis.

We have also looked carefully at the non-physical singularity that
is closer to the origin than the physical singularity. We find its
position to be at $u = -0.2853 \pm 0.0003$, and the exponents of
the susceptibilty, magnetisation and specific heat to be $\approx
1.03$, $-0.01$ and $1.01$ respectively. It is difficult not to suggest
that these are 1, 0 and 1 respectively, with presumably logarithmic
corrections. These values also satisfy the same scaling relation
$\alpha' + 2\beta + \gamma' = 2$ that the exponents at the critical point
satisfy, though the thermodynamic argument that leads to this result
at the critical point is not obviously
 applicable in this non-physical region.

For the high-temperature series, the critical temperature is the radius
of convergence of the series. It follows that the ratio method and its
variants can be used. Unbiased differential approximants, using the
method of analysis described in Guttmann (1987a) give $v_c^{2} =
0.0475185$,
$\alpha = 0.126$. Utilising the linear dependence between the estimates
of $v_c$ and $\alpha$ gives a biased estimate of 0.11 for $\alpha$ at the
correct $v_c^{2} = 0.047565$.

With a `correction-to-scaling' exponent close to 0.5, the ratio of succesive
terms in the specific heat expansion is expected to behave as
$	r_n = \mu[1 + (\alpha-1)/n + c/n^{3/2} + O(1/n^{2})] $
so estimators of $\alpha$ are given by the sequence
$       (r_n/\mu - 1)n + 1 = \alpha_1 + c/n^{1/2} + O(1/n)  $
where $\mu = v_c^{-2}$.
In Fig. 1  we show the plot of this sequence against $1/n^{1/2}$.
The series we have used is the expansion of ${C_H \over R}$ in powers of $v^2$,
just as was used by Sykes (1972), but with three further terms.
Linear
regression to the data gives $\alpha = 0.113 - 0.0637/n^{1/2} + 0.389/n$.
As we have shown in previous work, the method of differential approximants
can accurately predict the most significant digits of the next term in
the series (see e.g. Guttmann and Enting 1988a). In this way we
have estimated the coefficient of $v^{24}$ in the zero-field high-temperature
partition function to be $2.73376 * 10^{11}$, where only the last
quoted digit is uncertain. Using the additional term in the analysis
just described, we obtain an estimate for $\alpha$ of 0.111, with the
correction terms given by $-0.0511/n^{1/2} + 0.364/n$. This estimate of
$\alpha$ is impervious to the uncertainty in the estimated coefficient.
We conclude from this analysis that $\alpha = 0.110 \pm 0.005$.
The comparatively small amplitude of the
term proportional to $1/n^{1/2}$ helps us to understand why the series is
so well behaved.

Recently, Liu and Fisher (1990) studied the correction-to-scaling
amplitudes of the three-dimensional Ising model and argued
that the amplitudes of the correction-to-scaling term should be
negative.
Writing $C \approx A|t|^{-\alpha}[1 + a_\theta|t|^\theta + a_1 t + \dots]$
where $t = (T-T_c)/T_c$ and $\theta = \frac{1}{2}$ our result
transforms to $a_\theta \approx -0.040$, which has the predicted
sign.

 Biased first-order differential approximants give
$\alpha = 0.104 \pm 0.018$, where the analysis is precisely that described
in Guttmann (1987a). The error bars correspond to two standard deviations.
A Roskies-transformed Pad\'e analysis
was also performed for the high-temperature
series, paralleling the low-temperature investigation. The short series meant
that very few approximants were obtained, and so the results are not
particularly helpful (exponent estimates in the range $0.08$ to $0.3$ were
obtained).

\section{Concluding remarks}

The calculations presented here have shown that on the simple
cubic lattice the finite lattice method of series
expansion compares favourably with conventional expansion
methods. However there is not the dramatic difference that occurs
for many two-dimensional problems. Nevertheless, significant extensions
of both high- and low-temperature series have been obtained.

 Our exponent estimates
are all consistent with field theory exponent estimates. The two best
behaved series give $\beta = 0.329  \pm 0.009$ and $\alpha = 0.110 \pm 0.005$.
The estimates of $\alpha'$ and $\gamma'$, while consistent with field
theory estimates, had rather wide error bars, and so the agreement
with field theory was less convinicing. Our estimates of $\alpha$ and
$\beta$ are not consistent with the conjectured exact values of Mojumder.

After acceptance of this paper, we received a preprint from Vohwinkel (1992),
as
well as some additional series. Vohwinkel has used a version of the
shadow-lattice
method of Sykes, and has obtained six further low temperature series
coefficients.
Analysis of these longer series by the methods used in this work do not
significantly change our exponent estimates.

\bigskip\noindent

\ack{Financial support from the Australian Research
Council is acknowledged. We are grateful to M.F. Sykes
who confirmed some of our new series coefficients, and to Claus
Vohwinkel for preprints and series prior to publication. The series were
counted
primarily on an IBM 3090/400J, and we would like to than the Australian
Communications and Computing Institute for the provision of the facility,
and Glenn Wightwick and Jan Jager for their help in running such
demanding jobs.}

\references
\refjl{Adler J, Enting I G and Privman V 1983}{\JPA}{16}{1967--73}
\refjl{Bhanot G, Creutz M and Lacki J 1992}{Phys. Rev. Lett.}{69}{1841}
\refjl{Camp W J, Sauk D M, Van Dyke J P and Wortis M
1976}{Phys. Rev. B}{14}{3990}
\refjl{Enting I.G. 1978a}{Aust. J. Phys}{31}{515--22}
\refjl{\dash 1978b}{\JPA}{11}{563--68}
\refjl{\dash 1980a}{\JPA}{13}{3713--22}
\refjl{\dash 1980b}{\JPA}{13}{L133--6}
\refjl{\dash 1980c}{\JPA}{13}{L409--12}
\refjl{\dash  1980d}{\JPA}{13}{L279--84}
\refjl{\dash 1987a}{\JPA}{20}{1485--94}
\refjl{\dash 1987b}{\JPA}{20}{L917--21}
\refjl{Enting I G and Guttmann A J 1985}{\JPA}{18}{1807--17}
\refjl{\dash 1989}{\JPA}{22}{1371--84}
\refjl{\dash 1992}{\JPA}{25}{2791--807}
\refjl{Enting I G and Wu F Y 1982}{J. Statist. Phys}{28}{351--73}
\refjl{Gaunt D S and Sykes M F 1973}{\JPA}{6}{1517}
\refjl{Guttmann A J 1987a}{\JPA}{20}{1839--54}
\refjl{ \dash 1987b}{\JPA}{20}{1855--63}
\refbk{\dash 1989 Asymptotic Analysis of Power Series Expansions.
 In: C. Domb and J. Lebowitz (eds)}{Phase
Transitions and Critical Phenomena, Vol. 13}{(Academic, New York).}
\refjl{Guttmann A J and Enting I G 1988a}{\JPA}{21}{L165--72}
\refjl{\dash 1988b}{\JPA}{21}{L467--74}
\refjl{\dash 1990}{Nucl. Phys. B (Proc. Suppl.)}{17}{328--330}
\refjl{Hunter, C 1987}{SIAM J. of Appl. Math.}{47}{483}
\refjl{Kim D. and Enting I.G. 1979}{J. Comb. Theor}{26B}{327--36}
\refjl{Le Guillou J C  and Zinn-Justin J 1980}{Phys. Rev. B}%
{21}{3976}
\refjl{Liu A J and Fisher M E 1990}{J. Stat. Phys. }{58}{431}
\refjl{Mojumder M A  1991}{Mod. Phys. Lett.}{6}{2687--91}
\refbk{de Neef T 1975 }{ Some applications of series
expansions in magnetism.}{ Ph.D. Thesis. Technische Hogeschool
Eindhoven}
\refjl{de Neef T and Enting I G 1977}{\JPA}{10}{801-5}
\refjl{Nickel B G 1991}{Physica}{177}{189--96}
\refjl{Oitmaa  J, Hamer C J and Zheng W 1991}{\JPA}{24}{2863--67}
\refjl{Roskies R Z 1981}{Phys. Rev. B}{24}{5305}
\refjl{Sykes M F, Hunter D L, McKenzie D S and Heap B R 1972}%
{\JPA}{5}{667--73}
\refjl{Sykes M F, Gaunt D S, Essam  J W and Elliott C J 1973}%
{\JPA}{6}{1507}
\refbk{Vohwinkel C 1992}{preprint}{DESY, Hamburg}

\tables

$$\vcenter{
\halign{\hfil#\hfil&\quad\hfil#&\qquad\hfil#&\quad\hfil#&\qquad\hfil#\cr
\noalign{\hrule}
\noalign{\smallskip}
 Cross-section& $s$ &\multispan 2 \qquad Order of Potts (Ising)
series &  \hfil $R(d \times w,2)$\hfil\cr
 & & Low-$T$ & High-$T$ &\cr
\noalign{\smallskip}
\noalign{\hrule}
\noalign{\smallskip}
$1 \times 1 $ & 4  & 13 (6)   &  3 (2)  & 2 \cr
$1 \times 2 $ & 5  & 17 (8)  &  5 (4)  & 4 \cr
$2 \times 2 $ & 7  & 25 (12)  &  9 (8)   & 16 \cr
$2 \times 3 $ & 8  & 29 (14)  &  11 (10)   & 64  \cr
$3 \times 3 $ & 10  & 37 (18)  &  15 (14)  & 512 \cr
$3 \times 4 $ & 11  & 41 (20)  &  17 (16)  & 4096  \cr
$4 \times 4 $ & 13  & 49 (24)  &  21 (20)  & 65536 \cr
$4 \times 5 $ & 14  & 53 (26)  &  23 (22) & 1048576 \cr
$ 5 \times 5 $ & 16  & 61 (30)  & 27 (26)  &  33554432 \cr
\noalign{\smallskip}
\noalign{\hrule}
}}$$

\tabcaption{ Combinatorial factors determining the computational
complexity of
finite lattice series expansions on the simple cubic lattice.  Column 1
gives the various
possible cutoff points as specified by the largest cross-section that need
be considered
after making optimal use of cubic symmetry.  The second column is $s$, the
largest value of
the sum of length plus width plus depth for any of the cuboids used in the
expansion.  The
low-temperature series for the Potts model can be obtained to order
$z^{4s-3}$ and the
high-temperature series to $v^{2s-5}$.  Ising model series can be obtained
to $u^{2s-2}$ and $v^{2s-6}$. The final column, $R(.,2)$, gives the
number of vector elements required for the low-temperature Ising
expansion.}

$$\vcenter{
\halign{\hfil#\hfil&\quad\hfil#&\qquad\hfil#&\quad\hfil#&\qquad\hfil#\cr
\noalign{\hrule}
\noalign{\smallskip}
$n$   &  $\lambda_n$ & $m_n$   &  $c_n$  &  $a_n$ \cr
\noalign{\smallskip}
\noalign{\hrule}
\noalign{\smallskip}
 0   &    1    &    1   &    0    &   1  \cr
 3   &    1    &    -2   &   1     &   0  \cr
 4   &    0    &    0   &   0     &   3 \cr
 5   &    3    &    -12   &  12      &   0 \cr
 6   &    -3    &    14   &   -14     & 22 \cr
 7   &    15    &    -90   &  135      &  0 \cr
 8   &    -30    &    192  &  -276      &   192 \cr
 9   &    101    &   -792  &  1520     &   0 \cr
 10   &    -261    &  2148  & -4056       &   2046 \cr
 11   &    807    &  -7716 &  17778      &   0 \cr
 12   &  -2308    & 23262  &  -54392      &   24853 \cr
 13   &  7065      & -79512 &  213522      &   0 \cr
 14   &  -21171      & 252054 &  -700362      &   329334 \cr
 15   &   65337     & -846628 &  2601674   &   0 \cr
 16   &   -200934     & 2753520 & -8836812   &   4649601  \cr
 17   &   627249     & -9205800 & 31925046   &   0 \cr
 18   &   -1962034     & 30371124 & -110323056   &   68884356 \cr
 19   &    6192066    & -101585544 & 393008712   &   0 \cr
 20   &    -19610346    & 338095596 &  -1369533048  &   1059830112 \cr
 21   &    62482527    & -1133491188 & 4844047090   &   0 \cr
 22   &    -199807110    &  3794908752 & -16947396000   &   16809862992 \cr
 23   &    641837193    & -12758932158 &  59723296431  &   0 \cr
 24   &  -2068695927      & 42903505030 &  -209328634116  &   27337xxxxxxx \cr
 25   &   6691611633     & -144655483440 & 736260986208   &   0 \cr
 26   &  -21710041944      & 488092130664 &  -2582605180212   &    \cr
\noalign{\smallskip}
 \noalign{\hrule}
}}$$

\tabcaption{Coefficients in low-temperature expansions for $\Lambda_0$,
$M$ and $\chi$ and high-temperature expansion for $\Phi$,
defined by equations (9), (10), (11) and (17). The
incomplete coefficient, $a_{24}$,
was not derived using the finite lattice method, but represents
an extrapolation obtained using differential approximants.}

$$\vcenter{  
\halign{  
\hfil$#$\qquad&\hfil$#$\qquad&\hfil$#$\qquad&\hfil$#$\qquad&\hfil$#$\qquad&\
\hfil$#$\cr
N&[N-2/N]&[N-1/N]&[N/N]&[N+1/N]&[N+2/N]\cr
\noalign{\medskip}
5&0.4192&0.3590&0.2290&0.5032&0.3240\cr
6&0.3261&0.3346&0.3431&0.3437&0.3325\cr
7&0.4559&0.3437&0.3431&0.3332&0.3326\cr
8&0.3027&0.3388&0.3258&0.3342&0.3322\cr
9&0.3358&0.3419&0.3731&0.2900&0.3279\cr
10&0.3275&0.3186&0.3580&0.3215&0.3230\cr
11&0.3270&0.3318&0.3233&0.3210&0.3391\cr
12&0.3290&0.3274&0.2624&0.2113&0.3286\cr
13&0.3286&0.2154\cr}}$$
\bigskip\noindent

\tabcaption{Estimates of $\beta$ using Dlog Pad\'e
approximants to Roskies-transformed series.}

$$\vcenter{  
\halign{  
\hfil$#$\qquad&\hfil$#$\qquad&\hfil$#$\qquad&\hfil$#$\qquad&\hfil$#$\qquad&\
\hfil$#$\cr
N&[N-2/N]&[N-1/N]&[N/N]&[N+1/N]&[N+2/N]\cr
\noalign{\medskip}
5&1.1962&1.2342&1.2997&1.3836&1.1646\cr
6&1.4124&1.4975&1.3363&1.2835&1.1892\cr
7&1.4200&1.3068&1.3431&2.5047&1.1088\cr
8&1.7242&1.2416&1.2809&1.2551&1.2049\cr
9&1.2981&1.2655&1.2940&1.2087&1.2047\cr
10&2.7628&1.2466&1.2493&1.2767&1.2511\cr
11&1.2494&1.2461&1.2605\cr
12&1.2643\cr}}$$
\bigskip\noindent

\tabcaption{Estimates of $\gamma'$ using Dlog Pad\'e
approximants to Roskies-transformed series.}

$$\vcenter{  
\halign{  
\hfil$#$\qquad&\hfil$#$\qquad&\hfil$#$\qquad&\hfil$#$\qquad&\hfil$#$\qquad&\
\hfil$#$\cr
N&[N-2/N]&[N-1/N]&[N/N]&[N+1/N]&[N+2/N]\cr
\noalign{\medskip}
5&0.1850&0.1197&0.1102&0.1410&0.0829\cr
6&0.1078&0.1173&0.1236&0.1319&0.1265\cr
7&0.1224&0.1131&0.1284&0.1329&0.1109\cr
8&0.1260&0.1207&0.1239&0.1261&0.1364\cr
9&0.1234&0.1686&0.1226&0.1208&0.1658\cr
10&0.1221&0.1214&0.1227&-0.686&0.2232\cr
11&0.1221&0.1182&0.1743\cr
12&0.1375\cr}}$$

\tabcaption{Estimates of $\alpha'$ using Dlog Pad\'e
approximants to Roskies-transformed series.}

\figures
\figcaption{Ratio method estimates of $\alpha$, plotted
against $n^{1/2}$. The series used was ${C_H \over R}$, expanded
 in powers of $v^2$. The slope gives the amplitude
of the `correction-to-scaling' term and the intercept
gives the estimated value of $\alpha$ after taking
the correction-to-scaling into account.}

\end